\documentclass[prb,twocolumn,floatfix]{revtex4}
\usepackage{graphics}
\usepackage{epsfig}

\renewcommand{\in}{{\rm ab}}
\newcommand{\out}{{\rm c}}
\newcommand{\ex}{{\rm ex}}

\begin{document}

\title{Anisotropy of exchange stiffness and its effect on the properties of magnets}
\author{K. D. Belashchenko}
\affiliation{Department of Physics and Astronomy, University of Nebraska, Lincoln, Nebraska 68588}

\begin{abstract}
Using the spin-spiral formulation of the tight-binding linear muffin-tin orbital method,
the principal components of the exchange stiffness tensor are calculated for typical hard magnets
including tetragonal CoPt-type and hexagonal YCo$_5$ alloys. The exchange stiffness is strongly
anisotropic in all studied alloys.
This anisotropy makes the domain wall surface tension anisotropic.
Competition between this anisotropic surface tension and magnetostatic energy controls
the formation and dynamics of nanoscale domain structures in
hard magnets. Anisotropic domain wall bending is described in detail from the
general point of view and with application to cellular Sm-Co magnets. It is shown
that the repulsive cell-boundary pinning mechanism in these magnets is feasible
only due to the anisotropic exchange stiffness if suitably oriented initial
pinning centers are available. In polytwinned CoPt-type magnets the exchange stiffness
anisotropy controls the orientation of macrodomain wall segments. These segments may
reorient both statically during microstructural coarsening and dynamically during
the macrodomain wall splitting in external field. Reorientation of segments
may facilitate their pinning at antiphase boundaries.
\end{abstract}
\maketitle

\section{Introduction}

The formation and dynamics of magnetic domain structures
are commonly studied by micromagnetic
methods employing the phenomenological gradient expansion for
the `exchange term' in the free energy.~\cite{Kittel} For a crystal
of arbitrary symmetry this term may be generally written as
\begin{equation}
F_{\ex}=A_{\alpha\beta}\int\frac{\partial e_\gamma}{\partial r_\alpha}
\frac{\partial e_\gamma}{\partial r_\beta}\, d^3 r
\label{free}
\end{equation}
where
${\bf e}({\bf r})={\bf M}({\bf r})/M$ is the unit vector parallel
to magnetization, summation is assumed over repeated Cartesian
indices, $A_{\alpha\beta}=MD_{\alpha\beta}/4$, and $D_{\alpha\beta}$
is the spin-wave stiffness tensor which determines the long-wavelength
part of the magnon spectrum as
$\hbar\omega({\bf q})=D_{\alpha\beta}q_\alpha q_\beta$.
In a cubic crystal
$A_{\alpha\beta}=A\,\delta_{\alpha\beta}$ where $\delta_{\alpha\beta}$ is
the Kronecker symbol, and $A$ is commonly referred to as the exchange constant.
Below $A_{\alpha\beta}$ is referred to as the exchange stiffness tensor.

A cubic crystal may only have a fourth-order magnetocrystalline
anisotropy (MCA) in the spin-orbit coupling parameter $\xi$,
while a non-cubic crystal has MCA in the second order in $\xi$.
Since magnetic hardness generally requires high MCA, all known
hard magnets are non-cubic.
Many of them are uniaxial, so that $A_{\alpha\beta}$ has
two principal components, in-plane $A_{\in}$ and out-of-plane
$A_{\out}$.

The components of the exchange stiffness tensor may be found both
theoretically using non-collinear spin-polarized band structure
calculations, and experimentally, from the long-wavelength part of the magnon
dispersion spectrum. Nevertheless, to my knowledge, none of these
methods was applied to hard magnets, and the isotropic model
($A_{\out}=A_{\in}=A$) was explicitly used in all micromagnetic
calculations (see, e.g.,
Refs.~\onlinecite{Kron,Schrefl}), while $A$ is usually estimated from
the Curie temperature.~\cite{Skomski-book} This paper reports the
results of calculations of $A_{\in}$ and $A_{\out}$ for several
uniaxial hard magnets including CoPt, FePt,
FePd and the SmCo$_5$-like compound YCo$_5$. It turns out that
exchange stiffness anisotropy in hard magnets is typically quite large.

Strong anisotropy of exchange stiffness may significantly affect
the hysteretic properties of a magnet, because 
it translates into anisotropic domain wall surface tension:
\begin{equation}
\gamma_n=4(A_nK)^{1/2}
\label{DWST}
\end{equation}
where $K$ is the MCA constant, and $A_n$ is the exchange stiffness along the
direction normal to the domain wall. Therefore, the domain walls have a tendency to align
normal to the magnetically soft direction (the one with the lowest $A_n$).

On the other hand, according to the pole avoidance principle, the magnetostatic
contribution to the free energy prefers to eliminate the `magnetic charges'
$\rho=-\mathop{\rm div}{\bf M}$ localized on the domain walls by aligning
them parallel to the magnetization axis. For a magnet with easy-axis MCA,
if $A_{\in}<A_{\out}$ then the exchange term favors the
alignment of domain walls parallel to the easy axis, just as the
magnetostatic term. However, if $A_{\in}>A_{\out}$
as in all studied alloys,
then the exchange and magnetostatic terms favor
different (orthogonal) domain wall orientations. The relative importance
of these terms depends on the length scale and geometry of the domain
structure. As we will see below, the exchange term often dominates in hard magnets
with sufficiently fine domain structures. In this case common considerations
based on the pole avoidance principle are inapplicable, and various peculiarities
in the domain structure and its response to the external magnetic field should be expected.


The rest of the paper is organized as follows. Anisotropy of exchange
stiffness in several typical hard magnets is determined using
first-principles calculations in Section~\ref{calc}.
Temperature dependence of the exchange stiffness
anisotropy is discussed in Section~\ref{temp}.
The following sections describe the effects of exchange stiffness
anisotropy on the properties of the domain structure. Anisotropic domain
wall bending is considered in Section~\ref{bending} from the
general point of view. The effect of anisotropic domain wall bending
on the coercivity of cellular Sm-Co magnets is addressed in Section~\ref{smco}.
Section~\ref{copt} examines the effects of exchange
stiffness anisotropy on the structure of domain walls and coercivity of
polytwinned CoPt type magnets. Section~\ref{concl} concludes the paper.

\section{Calculation of exchange stiffness in C\lowercase{o}P\lowercase{t}
type and YC\lowercase{o}$_5$ magnets}\label{calc}

The values of $A_{\in}$, $A_{\out}$ were calculated using the
spin-spiral formulation~\cite{SS} of the tight-binding linear muffin-tin
orbiral (TB-LMTO) method within the atomic sphere approximation (ASA)
including the `combined correction' term.~\cite{Andersen}
The spin-spiral (``frozen magnon'') approach is much more
reliable compared to the calculation of exchange stiffness based on
a finite number of pair exchange parameters, because in practice the
corresponding sum in real space does not converge.

Local spin density approximation (LSDA) was used
with von Barth-Hedin exchange-correlation potential.~\cite{vBH}
The calculations were done for experimental
values~\cite{CoPtstr,Pearson,YCo5str} of lattice constants given
in Table I. The atomic sphere radii for both constituents
were taken to be equal to each other in CoPt, FePt and FePd.
In YCo$_5$ the radii were 3.548 a.u.
for Y and 2.627 a.u. for Co in both inequivalent positions.
The calculations were repeated for two setups with three
($l_{\rm max}=2$) and four ($l_{\rm max}=3$) basis functions
per atom, and care was taken to achieve convergence with the
Brillouin zone sampling. Spin-orbit coupling was neglected.

In the spin-spiral method the direction of the magnetic moment
at site $i$ depends on coordinates as
\begin{equation}
{\bf e}_i=(\sin\theta\cos{\bf qr}_i,\sin\theta\sin{\bf qr}_i,\cos\theta)
\label{spiral}
\end{equation}
where ${\bf q}$ is the wave vector and $\theta$ the amplitude
of the spin spiral. For a spin spiral with small $q$ at zero
temperature, according to (\ref{free}), we have
\begin{equation}
E/V=A_{\alpha\beta}\;q_\alpha q_\beta\,\sin^2\theta.
\label{EexSS}
\end{equation}
where $E$ and $V$ are the excess total energy referenced from the
ferromagnetic state and volume of the computational cell.

The results listed in Table~\ref{tab} show that the exchange stiffness
anisotropy is quite large in all studied alloys. Notably, $A_\in$ is everywhere
greater than $A_\out$ due to the predominantly in-plane
bonding between Fe or Co atoms. For comparison, the values of $A_\in$
and $A_\out$ were also calculated for hcp cobalt. Here the exchange
stiffness is almost isotropic as expected, and
its value of approximately 3.5$\times10^{-6}$~erg/cm
(or $D\simeq600$~meV$\cdot$\AA$^2$) is in good agreement with
experiment~\cite{BLSW} and with other calculations.~\cite{Cooke,Liu}
The exchange stiffness for both directions is larger in YCo$_5$ compared
to CoPt type alloys due to higher Co concentration.
$A_\in$ in YCo$_5$ is close to that in hcp Co.

\begin{table}[ht]
\caption{Lattice parameters and calculated values of in-plane and
out-of-plane exchange stiffness (units of 10$^{-6}$~erg/cm).
The value given before (after) the slash
was calculated with $l_{\rm max}=2$ ($l_{\rm max}=3$).}
\begin{tabular}{|l|c|c|c|c|c|}
\hline
&CoPt&FePt&FePd&YCo$_5$&Co\\
\hline
$a$, \AA&3.806&3.861&3.860&4.937&2.507\\
\hline
$c/a$&0.968&0.981&0.968&0.806&1.623\\
\hline
$A_{\in}$&1.70/1.58&1.10/0.87&1.89/1.78&3.97/3.89&3.55\\
\hline
$A_{\out}$&1.13/1.03&0.38/0.06&0.87/0.72&1.68/1.57&3.43\\
\hline
$\alpha=A_{\out}/A_{\in}$&0.66/0.65&0.34/0.07&0.46/0.40&0.42/0.40&0.97\\
\hline
\end{tabular}
\label{tab}
\end{table}


The out-of-plane exchange stiffness in FePt is unusually small and very sensitive
to the lattice parameter $c$. This magnetostructural effect may be quantified by
the value $W=dA_{\out}/d\ln c=26\cdot10^{-6}$~erg/cm.
Low value of $A_\out$ and high value of $W$ imply that moderate compression of the
order of 1--2\% along the $c$ axis may induce magnetic instability in FePt with
the formation of a spin wave in the $c$ direction. This conclusion agrees qualitatively
with the results of other studies suggesting that the layered antiferromagnetically
ordered (AFM) state (a special case of such spin wave) in FePt has lower energy compared
to the ferromagnetic (FM) state under moderate $c/a$ reduction~\cite{Zeng} or even at
experimental lattice parameters.~\cite{Brown} However, strong sensitivity of $A_\out$
to the basis set (see Table 1) indicates that ASA is too crude for the description of
magnetic energetics in FePt. On the other hand, the LSDA approximation also seems to
be insufficient, because adding any of the two types of gradient corrections~\cite{LMH,PW}
to the LSDA exchange-correlation potential notably tends to stabilize the FM phase.

Competition between different magnetic structures (including non-collinear ones) is
characteristic for fcc phases of iron~\cite{Nordstrom} and its alloys (Fe$_3$Pt is a known
Invar alloy), and hence it is not surprising for FePt. Indeed, while FePt is tetragonal,
its structure fully retains the topology of the fcc lattice if Fe and Pt sites are
considered equivalent.

From the practical point of view, structural sensitivity of $A_\out$ in FePt
suggests that the exchange stiffness anisotropy in this magnet may be controlled
using chemical pressure,~\cite{Zeng} appropriate doping or
off-stoichiometry. In view of the strong effect of this
anisotropy on the hysteretic properties (see below), this possibility may
prove useful in applications, such as the design of perpendicular magnetic
recording media.

\section{Temperature dependence}\label{temp}

It is obvious that exchange stiffness anisotropy $\alpha$ is an additional
parameter of micromagnetics which may have a strong effect on the coercivity
and other properties of magnets.
In this connection it is worth noting that in some materials $\alpha$
may strongly depend on temperature and doping. For example,
consider a layered magnet with atoms of type A in even layers and
type B in odd layers. Suppose that the exchange interaction
is strong for A--A pairs, negligible for B--B pairs, and small for
A--B pairs. This is a good approximation for all CoPt type magnets,
where A corresponds to the 3d metal, and B to Pt or Pd.
If we assume that the magnitudes of the magnetic moments $M_A$ and
$M_B$ do not depend on temperature (rigid local moments model),
in the mean field approximation the reduced magnetization of the B layer
$m_B=\langle{\bf M}_B\rangle/M_B$ is
\begin{equation}
m_B=f(2J_{BA}m_A/T)
\end{equation}
where $m_A=\langle{\bf M}_A\rangle/M_A$, $f(x)=\coth(x)-1/x$, and
$J_{BA}=\sum_{j}J_{ij}$ where site $i$ is within the B layer and $j$
runs over A sites.
Since $J_{BA}$, as we assumed, is much smaller than $J_{AA}$ (defined
with both $i$ and $j$ in the A layer), there is
a wide range of temperatures where the alloy is still ferromagnetic,
but $J_{BA}/T\lesssim1$. In this region $f(x)\approx x/3$,
and, supposing that the Curie temperature $T_c$ is almost entirely
determined by A--A interactions ($T_c\approx2J_{AA}/3$), we obtain
\begin{equation}
\frac{m_B}{m_A}\approx\frac{J_{BA}}{J_{AA}}\frac{T_c}{T}\quad,
\label{T-dep}
\end{equation}
that is, the ratio of magnetizations of B and A sublattices is
inversely proportional to temperature.
Obviously, the relative contribution of A--B pairs to $A_\in$ and $A_\out$
in the mean-field approximation follows the same law.
At the same time, due to the layered structure the A--B pairs may give an
important contribution to $A_\out$ at $T=0$ (according to the calculation,
this is the case in FePt). Eq.~(\ref{T-dep}) implies that close to $T_c$
this contribution is reduced by a factor ${J_{BA}}/{J_{AA}}$, and hence
$\alpha$ is essentially determined only by exchange interaction in A--A pairs.

In addition, in magnets like FePt the induced magnetic moments of Pt atoms
should be more easily destroyed by thermal excitations compared to the
self-induced, well-localized Fe moments, and the temperature dependence
of $\alpha$ should be even more pronounced.

On the other hand, non-magnetic impurities in
a layered system (e.g. Cu in SmCo$_5$) may energetically prefer
some specific layers. At high concentration of such
impurities the interlayer exchange coupling will be strongly reduced,
again decreasing $\alpha$ and also enhancing its temperature dependence.

These effects provide an interesting mechanism for the dependence of
magnetic properties on temperature and doping due to the increased
domain wall bending. This relationship may be important in Sm-Co
type magnets where domain walls are heavily bent, as discussed in 
Section~\ref{smco}.

\section{Anisotropic domain wall bending}\label{bending}

Bending of pinned domain walls in external magnetic field was invoked
by many authors to describe certain aspects of coercivity and hysteresis in
magnets.~\cite{Kersten,Globus,Zijlstra,HKcurv,Durst,Skomski}
This bending, which generally manifests itself in the initial magnetic
susceptibility, was incorporated in the Globus model to describe the
hysteresis loop of granular magnets.~\cite{Globus}
Pinning of bending domain walls on an array of defects was also considered
by a number of authors, see Ref.~\onlinecite{HKcurv} and references therein.

In real magnets domain wall bending may play a more subtle role in the
magnetization reversal. For some microstructures bending of domain walls may
facilitate their pinning by increasing the area of contact with pinning
centers. In particular, this mechanism was discussed in the studies of
coercivity and hysteresis loop in SmCo$_5$ powders~\cite{Zijlstra}
and in cellular Sm--Co magnets.~\cite{Durst,Skomski} 

In previous treatments of domain wall bending the exchange stiffness was
assumed to be isotropic. Here we will discuss the effect of
exchange stiffness anisotropy on the domain wall bending.

First, we will study domain wall bending neglecting the associated
stray fields. As we will see below, this approximation is valid when
the characteristic flux closure length is sufficiently small.

To get a general feeling of the problem of domain wall bending, it
is useful to invoke a direct analogy between domain walls and foam bubbles.
Indeed, in the presence of external magnetic field ${\bf H}$ the 
domain walls in a uniaxial magnet experience constant pressure of magnitude
$2\mathbf{MH}$ directed away from the regions where $\mathbf{HM}$ is positive.
On the other hand, if the exchange stiffness tensor
is isotropic, the free energy of a domain wall is simply proportional to
its total area, just as that of a foam membrane.

This analogy allows one
to guess the equilibrium configurations of domain walls pinned by certain
symmetric pinning sites. For example, a domain wall pinned by a circular defect
should obviously have the form of a sphere segment with radius $R$ 
related to the external field as $R=\gamma(\mathbf{MH})^{-1}$ (the Laplace pressure
$2\gamma/R$ of the curved domain wall compensates the applied
pressure $2\mathbf{MH}$). This result was obtained in Ref.~\onlinecite{Skomski} where
the sphere segment was used as a variational trial function.
A domain wall pinned at two parallel straight lines assumes cylindrical shape
with twice as smaller radius; this solution was
discussed for a domain wall in a thin ferroelectric film.~\cite{DWferroel}

If exchange stiffness is anisotropic, the analogy with the foam membrane
no longer holds, because the domain wall surface tension is also anisotropic.
Denoting the angle between the normal to the domain wall and the
magnetization axis as $\phi$, the surface tension of the
domain wall is given by (\ref{DWST}) with
\begin{equation}
A_n=A_{\in}\sin^2\phi+A_{\out}\cos^2\phi
\label{An}
\end{equation}

Consider a domain wall pinned at a closed curve $B$ 
with a typical size $R_0$ lying in the $x=0$ plane, choosing the $z$ axis
parallel to the easy magnetization axis. 
The shape of the domain wall is defined by a function $\xi=\xi(y,z)$ with
$\xi=0$ at the boundary $B$. The total free energy is easily shown to be
(we still neglect the stray fields and assume for simplicity that
$\mathbf{H}$ is parallel to the $z$ axis):
\begin{equation}
F=\int\left[\gamma_\in\sqrt{1+(\partial_y\xi)^2+\alpha(\partial_z\xi)^2}
-2MH\xi\right]dydz
\label{F-DW}
\end{equation}
where $\gamma_\in$ is the surface tension of a domain wall
parallel to the $z$ axis, the integral is taken over the area bounded by $B$, and
$\partial_y\equiv\partial/\partial_y$, etc. In the isotropic
case Eq.~(\ref{F-DW}) reduces to the expressions of
Refs.~\onlinecite{Skomski,DWferroel}. 
In zero field the equilibrium shape of the domain wall is just $\xi=0$, while
at $H\ne0$ it satisfies the Euler-Lagrange equation corresponding to the
variational problem $\delta F=0$:
\begin{equation}
\left(\partial_yR^{-1}\partial_y
+\alpha\partial_zR^{-1}\partial_z\right)\xi
+R_\in^{-1}=0\;,
\label{Euler}
\end{equation}
where $R$ is the square root from Eq.~(\ref{F-DW}), and $R_\in^{-1}=2MH/\gamma_\in$
is the domain wall curvature for ``in-plane bending'' (i.e. that with
$\partial_z\xi\equiv0$). The role of exchange stiffness anisotropy is seen most clearly
for the case when the external field is weak, $H\ll\gamma(MR_0)^{-1}$, and $\xi\ll R_0$.
In this case we obtain to first order in $\xi$:
\begin{equation}
\left(\partial^2_y
+\alpha\partial^2_z\right)\xi+R_\in^{-1}=0
\label{Euler1}
\end{equation}

\begin{figure}
\epsfig{file=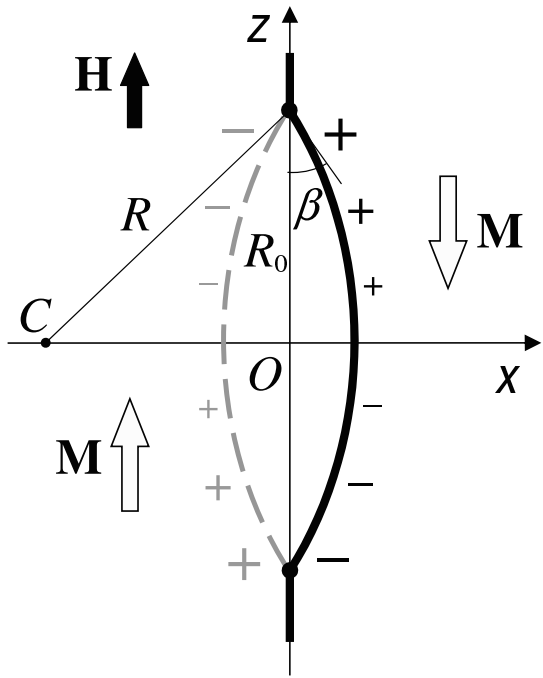,width=0.4\textwidth,clip}
\caption{Cylindrical out-of-plane bending of a domain wall (thick solid line) pinned
by two line defects located at $x=0$, $z=\pm R_0$. $C$ is the axis, and $R$ the radius
of the cylindrical domain wall. $\beta$ is the domain wall deflection
angle at the defect. The gray dashed line shows the fictitious domain wall used in
the calculation of the magnetostatic energy. The size of plus and minus symbols
schematically shows the magnetic charge density on the real and fictitious walls.
}
\label{lens}
\end{figure}

Eq.~(\ref{Euler1}) shows that the domain wall bends more easily ``out of plane''
($\partial_y\xi=0$) if $\alpha<1$ and ``in plane'' ($\partial_z\xi=0$) if $\alpha>1$.
Indeed, if the boundary $B$ is formed by two straight segments
parallel to the $z$ direction as in Ref.~\onlinecite{DWferroel}
(e.g., pinholes in a film) at distance $L=2R_0$ from each other,
then $\xi$ does not depend on $z$, and from (\ref{Euler1}) we obtain
the angle of domain wall deflection at the pinning site:
$\beta=2MHR_0/\gamma_\in$. On the other hand, if the boundary $B$
is formed by straight lines parallel to $y$ direction (e.g., 
scratches on a film surface), then $\xi$ does not depend on $y$,
and the deflection angle is $\alpha^{-1}$ times larger. The latter
configuration is shown in Fig.~\ref{lens}.

The practical implication of this result is that the efficiency
of pinning centers in a magnet with anisotropic exchange stiffness depends
on their orientation. In $\alpha<1$ case typical for hard magnets the domain
walls bend more easily when pinned by defects that are normal to the
magnetization axis.

The exact shape of the domain wall in external field may be found
from the solution of the Euler-Lagrange equation without the
assumption $\xi\ll R_0$. This yields the circular cylinder
segment~\cite{DWferroel} of radius $R_\in$ for in-plane bending,
and the elliptic cylinder segment for out-of-plane bending:
\begin{equation}
(\xi-\xi_0)^2+\alpha^{-1}z^2=R_\in^2
\end{equation}
where $\xi^2_0=R_\in^2-R_0^2/\alpha$.

Now, let us clarify the role of magnetostatic (stray)
fields. If the domain wall is parallel to the magnetization axis $z$,
it has no net magnetic charge. Any deviation from
this alignment produces magnetic charge on the wall with surface density
$\sigma=2M\cos\phi$, where $\phi$ is defined exactly as in Eq.~(\ref{An}).
Obviously, in-plane domain wall bending does not induce any charges on the wall,
and we should only be concerned about stray fields when we are dealing
with out-of-plane curvature.

To estimate when stray fields may notably affect out-of-plane domain wall
bending, we have to compare the magnetostatic energy $\delta\epsilon_m$
generated by domain wall charging to the excess surface free energy
$\delta\epsilon_s$ associated with this bending (both energies are defined
per unit length in the $y$ direction). We assume that the domain
wall is pinned by two line defects parallel to the $y$ axis and displaced
from each other by a distance $L=2R_0$ along the $z$ axis, as shown in Fig.~\ref{lens}.
We will find $\delta\epsilon_m$ and $\delta\epsilon_s$
for a domain wall bent in a weak field. As follows from Eq.~(\ref{Euler1}), in this case
the domain wall is shaped as a segment of a circular cylinder of large radius
$R=\alpha R_\in$. Substituting this solution in Eq.~(\ref{F-DW}) we obtain
$\delta\epsilon_s\approx\frac13\alpha\gamma_\in R_0^3/R^2$.

The magnetostatic energy is given by the integral
\begin{equation}
\delta\epsilon_m=-\frac12\int\mathbf{MH}_mdxdz
\label{dem}
\end{equation}
where $\mathbf{H}_m$ is the stray
field generated by the surface charges on the domain wall.
The distribution of these charges is antisymmetric with respect to the $z=0$ plane
(see Fig.~\ref{lens}), and the stray field obviously falls off at the length scale of $R_0$.
The total positive charge per unit length of the wall is of the order
$\rho_+\approx MR_0^2/R$ (assuming $R\gg R_0$). Since the magnetization is reversed at
the domain wall, there is a strong cancellation in the integral (\ref{dem}).
Indeed, using the superposition principle, let us add a fictitious domain wall which
is a mirror image of the real domain wall with respect to the $x=0$ plane.
This fictitious wall is shown in Fig.~\ref{lens} by the gray dashed line.
Mirror reflection also reverses the sign of the magnetic charges. For symmetry
considerations, the contribution to the integral (\ref{dem}) from \emph{outside}
of the lens-shaped area between the real and fictitious walls doubles when the fictitious
charges are added. At the same time, this contribution is negligibly small, because
the two walls form a thin capacitor, and the field is confined to its interior.
Therefore, only this interior region of cross-section $\frac43R_0^3/R$
contributes to $\delta\epsilon_m$. The stray field in this area is of the order
$2\rho_+/R_0$ (now we should take the field only from the real domain wall),
and from (\ref{dem}) we find $\delta\epsilon_m\approx\frac43M^2R_0^4/R^2$.
Thus, we obtain $\delta\epsilon_m/\delta\epsilon_s=\nu M^2R_0/(\alpha\gamma_\in)$
where the form-factor $\nu\approx4$. This relation also holds when the curvature
is not small ($R\sim R_0$), but $\nu$ should be somewhat different.

The relative importance of the magnetostatic energy increases linearly with $R_0$.
Using the relation $\gamma_\in=4K\delta_\in/\pi$ where $\delta_\in=\pi(A_\in/K)^{1/2}$ is the
in-plane domain wall width, we find that $\delta\epsilon_m$
overcomes $\delta\epsilon_s$ at $L\sim l_\mathrm{cr}=4\alpha\delta_\in/\eta$,
where $\eta=2\pi M^2/K$ is the dimensionless magnetostatic parameter
(in hard magnets $\eta$ is small, e.g. in Sm$_2$Co$_{17}$, CoPt and FePt it is
close to 0.1).
Thus, at $L\lesssim l_\mathrm{cr}$ the external field works mainly against the domain
wall surface tension, and its anisotropy is reflected in the domain wall bending
according to Eq.~(\ref{Euler}). At $L\gtrsim l_\mathrm{cr}$ the external field works
mainly against magnetostatic forces, which makes the contribution of surface tension
(together with its anisotropy) unimportant. By minimizing the total free energy given
by Eq.~(\ref{F-DW}) with $\delta\epsilon_m$ added, we find the 
deflection angle taking into account all energy terms: 
$\beta\simeq\beta_0L/(l_\mathrm{cr}+L)$ where $\beta_0=H/2M$. Note, however, that
this expression is approximate, because the equilibrium shape of the domain wall
at $L\gtrsim l_\mathrm{cr}$ is no more a cylindrical segment, and because we derived
it assuming $R\gg R_0$. At $L\ll l_\mathrm{cr}$ this result coincides with $\beta$ found above neglecting
the stray fields. However, as $L$ is increased beyond $l_\mathrm{cr}$, $\beta$
approaches its asymptotic limit $\beta_0$ and stops changing. 

In general, the parameter $l_\mathrm{cr}$ appears in all problems when magnetostatic
interaction competes with the domain wall energy. In a specific geometry,
the characteristic \emph{flux closure length} $l_f$ should be
compared with this parameter. At $l_f\gg l_\mathrm{cr}$ the magnetostatic 
interaction dominates; at $l_f\ll l_\mathrm{cr}$ it may be neglected.
In particular, the crossover length $l_\mathrm{cr}$ has the order of
the material parameter $\lambda$ of the magnetic bubble domain
theory (see, e.g., Ref.~\onlinecite{Bubbles}).
Together with $\eta$, this parameter also 
controls the structure of charged domain walls in thin films.~\cite{Hubert}
The critical single-domain size of a particle $R_{\rm sd}$
is also proportional to $l_\mathrm{cr}$;
for example, for a sphere with isotropic exchange stiffness
$R_{\rm sd}\approx5.6\delta/\eta$.~\cite{Skomski-book}

\section{Cellular S\lowercase{m}-C\lowercase{o} magnets}\label{smco}

Magnetization reversal in cellular Sm--Co magnets involves heavy bending
of domain walls. Since the anisotropy of exchange stiffness strongly
affects the ability of domain walls to bend, it is likely to play an
important role in the development of coercivity. In this section we
will focus on the model of repulsive domain wall pinning at the cell
boundaries and show that this mechanism can not be realized unless
the exchange stiffness is strongly anisotropic, and specifically oriented
line defects (such as those provided by the platelet phase) are available
for domain wall pinning \emph{in addition} to the cell boundaries.

Precipitation-hardened magnets based on an appropriately doped
Sm$_2$Co$_{17}$--SmCo$_5$ system develop outstanding magnetic hardness
in a wide range of temperatures. It is 
associated with the formation of a cellular microstructure
where rhomboid Sm$_2$Co$_{17}$-based (2:17) cells are surrounded by
SmCo$_5$-based (1:5) boundary phase~\cite{LM,FS,Del-Tang}
and is usually explained by domain wall pinning at the cell boundaries.~\cite{LM} Some
other mechanisms of coercivity were also suggested.~\cite{Tang02,Laughlin,Katter}

Whether pinning at the cell boundaries is attractive or repulsive depends on the
magnetic properties of 2:17 and 1:5 phases for the given (doped) system.
Recent experiments of Kronm\"uller and Goll~\cite{Kr-Goll} support the
hypothesis that pinning is repulsive at room temperature but
attractive at high temperatures, which also explains the anomalous temperature
dependence of coercivity (see Ref.~\onlinecite{Tang02} and references therein).

The estimated unpinning field for the cell boundaries (in the \emph{plane-parallel}
configuration) agrees with the experimentally observed coercivity.~\cite{Kr-Goll}
However, this does not fully explain
high coercivity, because magnetization reversal always takes the path
of lowest energetic barriers. It is not sufficient for a high barrier to
be present; it is necessary that there be no way around it. The domain walls might move
parallel to the hexagonal axis and never align parallel to the cell boundaries.
In order to be pressed against the cell boundaries, the domain walls must bend
in the external field. Skomski~\cite{Skomski} estimated the deflection angle $\beta$
of a domain wall at a pinning site at 56$^{\rm o}$ assuming $H=0.8$~T, cell size
$L=80$~nm, spherical bending, and isotropic exchange stiffness.

However, this value of the external field appears to be too high for
this estimate. Indeed, domain wall bending requires pinning at some `seed defects'
other than the cell boundaries.~\cite{similarconcept} In order to estimate the unpinning
field for these defects, we note that the coercivity of samples that
had not been subjected to slow cooling is only of the order of 0.1~T.~\cite{Kr-Goll}
The role of slow cooling is likely to promote the formation of the 1:5 phase with
segregated copper.~\cite{Kr-Goll} Assuming that the properties of the
2:17 cells are essentially unchanged during the slow cooling, we may take the value
of 0.1~T as an estimate of the unpinning field for the seed defects.
The same value corresponds to domain wall pinning at the vertices of the cells.~\cite{Skomski}
Thus, we obtain $\beta\approx6^\circ$ which is clearly insufficient
to press the domain wall against the cell boundary.

The above estimate assumes spherical bending and isotropic
exchange stiffness. In reality, it is reasonable to assume that the values
of exchange stiffness anisotropy in the 2:17 phases of the Sm-Co and Y-Co
systems are very close because exchange coupling is dominated by Co-Co pairs.
Indeed, the Curie temperatures of all R$_2$Co$_{17}$ phases (where R is a rare-earth
atom or yttrium) are almost identical,~\cite{Skomski-book} while the 2:17 phase
may be obtained from the 1:5 phase simply by a replacement of every third 
samarium atom by a Co$_2$ dumbbell.
Thus, we assume that the factor $\alpha$ in pure
Sm$_2$Co$_{17}$ is close to its value of 0.4 obtained for YCo$_5$ at $T=0$ (Table~1).
How does this affect the estimate of $\beta$?
As follows from Eq.~(\ref{Euler1}), exchange stiffness
anisotropy strongly facilitates domain wall bending only if it is pinned by
line defects normal to the magnetization axis (as in Fig.~\ref{lens}).
In this case the deflection angle $\beta$ contains an additional factor of $\alpha^{-1}$
due to exchange stiffness anisotropy and a factor of 2 due to the fact that
bending is cylindric instead of spheric. This brings $\beta$ to 30$^\circ$, neglecting
the effect of stray fields.

The magnetostatic term in the total energy of a bent domain wall is notable, although
not yet dominating for the cell size $L=80$~nm. Indeed, taking $A_\in=4\times10^{-6}$~erg/cm
and $\alpha=0.4$ as found above for YCo$_5$, $K=3\times10^7$~erg/cm$^3$ (Ref.~\onlinecite{Kr-Goll}),
and $M=950$~emu/cm$^3$ we obtain $\gamma_\in=44$~erg/cm$^2$, $\delta_\in=12$~nm, $\eta=0.15$,
and $l_\mathrm{cr}\approx 130$~nm. Using the approximate formula from Section~\ref{bending},
we arrive at the final estimate of $\beta\approx18^\circ$.

Let us summarize the results obtained above.
For the typical cell size of 80~nm the external field of 0.1~T (corresponding to
the unpinning of uncurved walls) may induce the domain wall deflection of about $18^\circ$
if the exchange stiffness anisotropy is large and if the domain walls may be initially
pinned by line defects normal to the magnetization axis. If either of these two conditions
is not met, the domain walls only deflect by about $6^\circ$ before they are
unpinned from the initial pinning centers. 

According to these estimates, the deflection angle does not reach the typical
cell-boundary inclination of $30^\circ$. Although our assumptions may be loosened up
to some extent (for example, allowing for a somewhat larger unpinning field for uncurved
walls), it seems clear that the repulsive cell-boundary pinning mechanism of coercivity
may only be realized in Sm--Co magnets under a very favorable set of circumstances.
In particular, it requires the presence of initial pinning defects normal to the $z$ axis.

High coercivity develops in Sm-Co magnets \emph{only} when they are doped with
zirconium~\cite{Zr-doping} which promotes the formation of thin lamellae normal
to the hexagonal axis of the crystal. The intersections of these lamellae with cell
boundaries have the ``right'' orientation needed for high coercivity in the repulsive
pinning case. Therefore, \emph{if} high coercivity of cellular Sm-Co magnets is due
to repulsive pinning at the cell boundaries, it is likely that the lamellar phase
provides the initial pinning centers for domain wall bending.
This conclusion does not contradict the observation~\cite{Kr-Goll}
that in the low-coercivity state obtained after annealing at 800$^\circ$C
the microstructure and the platelet phase are fully developed.
Indeed, the platelet phase simply provides initial pinning
sites which may be useful only in the presence of strongly pinning cell
boundaries developing only after the slow cooling. However, the microscopic
origin of pinning at the initial pinning sites is yet to be determined.

\section{Slanting and `smart pinning' of domain walls in
C\lowercase{o}P\lowercase{t}-type magnets}\label{copt}

In this section we will explore the effects of exchange stiffness anisotropy
on the structure of domain walls and magnetization reversal in polytwinned
CoPt-type magnets.
The microstructure of these magnets consists of regular stacks of
L1$_0$-ordered domains ($c$-domains). In each stack the $c$-domains are
separated by parallel twin boundaries in one of the \{110\}
planes.~\cite{Kandaurova-JMMM,Leroux91,Soffa2000,Khach,BPSV}
There is always a high density of antiphase boundaries within
the $c$-domains.~\cite{Leroux91,Soffa2000,Khach,BPSV,Shur-CoPt}

Usually the $c$-domain
thickness $d$ is large compared to the domain wall width $\delta\sim5$~nm,
and each $c$-domain may be regarded as an individual magnetic domain with
intrinsic 90$^\circ$ domain walls at the twin boundaries.~\cite{Kandaurova-JMMM}
The dynamic domain structure is formed by macrodomain
walls~\cite{Kandaurova-JMMM,Soffa2000} crossing many twin boundaries in a stack.
These walls are split at the twin boundaries, and their
segments are coupled only by relatively weak magnetostatic
forces.~\cite{BA-jap}

Below we study the effects of exchange stiffness anisotropy on the properties of
macrodomain walls. We will describe the orientation of macrodomain wall segments,
the energetical preference of different global macrodomain wall orientations,
the rotation of segments during macrodomain wall splitting in external field,
and the relation of these properties with coercivity.

\begin{figure}
\epsfig{file=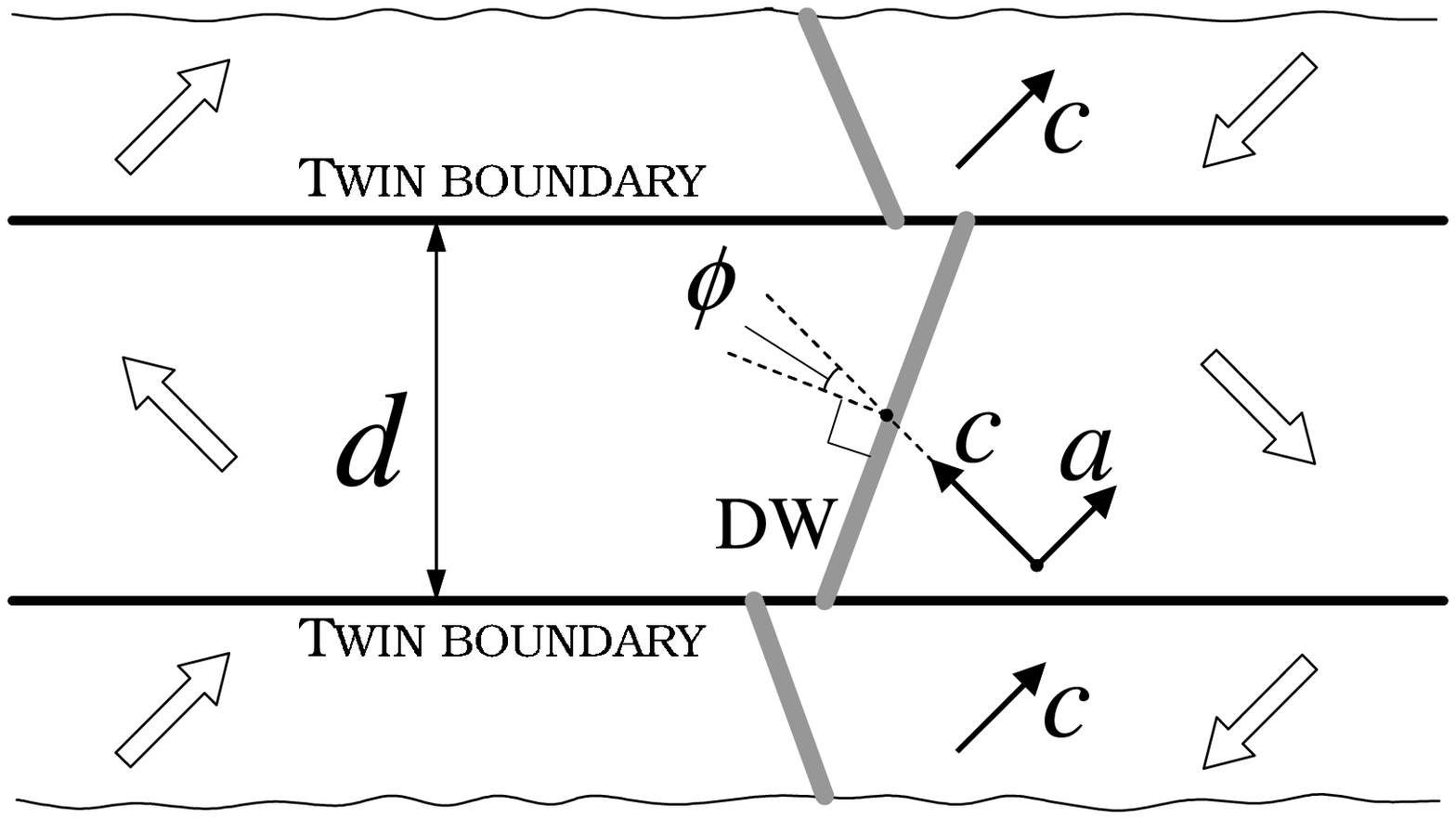,width=0.45\textwidth,clip}
\caption{Macrodomain wall in the polytwinned stack (three $c$-domains shown).
The central segment of the wall is marked DW. Large empty arrows show the 
of magnetization. Arrows marked `$c$' show the tetragonal axes in each $c$-domain;
arrow marked `$a$' shows one of the other two four-fold axes of the parent fcc lattice.
If $x$ and $y$ coordinates are assigned to $a$ and $c$ axes of the central $c$-domain,
the macrodomain wall has a global (1$\bar{1}0$) orientation.}
\label{slant}
\end{figure}

Consider a (1$\overline{1}0$) oriented macrodomain wall shown schematically in Fig.~\ref{slant}.
As it is shown by large empty arrows, the magnetization in each $c$-domain is parallel
to the easy $c$ axis, and it is reversed at the macrodomain wall.
The exchange stiffness $A_n$ is given by Eq.~(\ref{An}) where $\phi$, as shown in
Fig.~\ref{slant}, is now the angle between the normal to the domain wall segment and the
tetragonal axis $c$. Without magnetostatic interaction, the surface tension of the
macrodomain wall (e.g., the free energy of the domain wall segment per unit
normal cross-section) is $4(A_nK)^{1/2}\,[\cos(\frac{\pi}{4}-\phi)]^{-1}$.
Minimizing over $\phi$ we obtain
\begin{equation}
\tan\phi=\alpha
\label{orient}
\end{equation}
where, as above, $\alpha=A_\out/A_\in$. This result is analogous to a similar expression
obtained for an antiphase boundary.~\cite{Vaks-APB} We see that for
$A_\out=A_\in$ (isotropic exchange stiffness) $\phi=\pi/4$ as it should be~---
the domain wall has no preferential orientation and simply minimizes
its area by aligning perpendicular to the twin boundaries. For $A_\out\neq A_\in$ the domain
wall segments slant so as to decrease the surface tension.

Segments of a (1$\overline{1}0$) macrodomain wall carry magnetic charges of
alternating signs,~\cite{BA-jap} and the flux closure length $l_f$ is
obviously of the order of the $c$-domain thickness $d$. 
Therefore, at $d\gtrsim l_\mathrm{cr}$ the magnetostatic interaction dominates,
and the segments align parallel to the tetragonal axis to get rid of
the magnetic charge, while at $d\ll l_\mathrm{cr}$
their orientation is determined by Eq.~(\ref{orient}). Here $l_\mathrm{cr}$ is
proportional to $\delta/\eta$, but the form-factor is different from that
found in Section~\ref{bending} for domain wall bending.
In CoPt and FePt $l_\mathrm{cr}$ is of the order of 50~nm.
Thus, while $d$ increases during the microstructural coarsening,
the domain wall segments gradually slant from the angle given by
(\ref{orient}) towards the tetragonal axis.~\cite{Bloch-Neel}

Contrary to the (1$\overline{1}0$) oriented macrodomain wall, the segments of a (001)
oriented one are perpendicular to the twin boundaries at any $d$, because
this minimizes both the surface energy and the magnetostatic energy.
At $d\gtrsim l_\mathrm{cr}$ the (001) macrodomain wall has lower energy than
the (1$\overline{1}0$) one because its segments do not carry any magnetic charge.
However, at $d\ll l_\mathrm{cr}$ the preferential global orientation is
determined by exchange stiffness anisotropy.
The surface tension for (001) and (1$\overline{1}0$) macrodomain walls (per
unit normal cross-section)
is $\gamma_\in$ and $\gamma_\in\,[2\alpha/(1+\alpha)]^{1/2}$, respectively.
Therefore, at $\alpha > 1$ the (001) orientation is favorable, while in the
typical case $\alpha < 1$ the (1$\overline{1}0$) orientation is favorable.
This explains why macrodomain walls observed in FePt crystals have \{110\}
orientation~\cite{Kandaurova-JMMM} while those in FePd with much larger
$\eta$ and hence smaller $l_\mathrm{cr}$ have \{100\} orientation.~\cite{Zhang92}



High coercivity of CoPt-type magnets is likely due to the combination 
of macrodomain wall splitting and pinning of their segments at antiphase
boundaries.~\cite{BA-prb,BA-jmm}
The highest possible coercivity is achieved when antiphase boundaries are planar,
and domain wall segments are parallel to them.~\cite{DW-pinning}
The antiphase boundaries in CoPt-type magnets often have a preferential
crystallographic orientation, which is clear from theoretical
considerations,~\cite{Vaks-APB} and observed experimentally for
CoPt.~\cite{Leroux91} Typically they slant towards the
tetragonal axis, \emph{contrary} to the domain wall segments which slant
away from the tetragonal axis at $A_\out<A_\in$ and $d\ll l_\mathrm{cr}$.
This means that while $d$ increases during the microstructural coarsening, at some
point the domain wall segments will become parallel to the preferential orientation
of antiphase boundaries producing a maximum in the coercivity.

Interestingly, a similar segment rotation may occur dynamically.
If the external field above the splitting threshold~\cite{BA-prb} is applied
parallel to the twin boundaries, the macrodomain wall is split in two
``partial macrodomain walls'' which are driven apart from each other.
One partial wall is composed of segments in all odd $c$-domains, and the
other of segments in even $c$-domains.
Suppose that $d\ll l_\mathrm{cr}$ so that the orientation of
segments in the (1$\overline{1}0$) macrodomain wall is
given by Eq.~(\ref{orient}).
Since these segments carry magnetic charges of $\pm2M\cos\phi$ per unit
area, the two partial macrodomain walls carry equal charge
densities of opposite sign. The corresponding stray field makes
an additional contribution $\Delta E$ to the magnetostatic energy
proportional to the distance $L$ between the partial macrodomain walls.
If the angle $\phi$ were fixed, at $L\gtrsim l_\mathrm{cr}$ this
positive contribution would dominate over the surface energy of the segments
(now the flux closure length $l_f$ is clearly $L$).
Therefore, as the two partial macrodomain walls move apart ($L$ is increased to
$l_\mathrm{cr}$), their segments gradually rotate toward the tetragonal axis
to get rid of the magnetic charge.


As a result, at some $L$ the segments of a splitting macrodomain wall become parallel
to antiphase boundaries, just as in the case of increasing $d$ discussed above. 
If the typical distance between antiphase boundaries is smaller than $l_\mathrm{cr}$
(the scale of $L$ where domain wall segment rotation occurs), the segments will be
pinned by antiphase boundaries at the plain-parallel configuration.
In this scenario the coercivity achieves its highest possible value for suitably
oriented polytwinned stacks at any $d\ll l_\mathrm{cr}$, i.e. at relatively early
stages of coarsening. This mechanism may play an important role in real CoPt-type
magnets developing high coercivity just at these early stages.


\section{Conclusion}\label{concl}

Using the spin-spiral version of the TB-LMTO method,
the in-plane and out-of-plane principal components
of the exchange stiffness tensor were calculated for several
typical hard magnets. The results show that this tensor
usually has a considerable anisotropy. The out-of-plane
component is smaller than the in-plane one in all studied
hard magnets except pure hcp Co where exchange stiffness
is isotropic. The anisotropy is especially high in FePt.
In certain materials with intrinsically non-magnetic layers
(CoPt, FePt, etc.) the anisotropy of exchange stiffness may strongly
increase at finite temperatures or with suitable non-magnetic doping.

Anisotropy of exchange stiffness may have a strong effect on the
orientation of domain walls and on their resistance
to bending, and hence on the hysteretic properties of the magnet.
These effects may be expected whenever the typical flux
closure length associated with the stray fields does not exceed
the crossover length $l_\mathrm{cr}\propto\delta/\eta$
(see Section~\ref{bending}).


Low out-of-plane exchange stiffness facilitates out-of-plane domain
wall bending. This effect is crucial for the development of coercivity
in cellular Sm-Co magnets in the repulsive cell-boundary pinning regime.
This regime may be realized only in the presence of linear pinning defects
normal to the magnetization axis (such as the intersections of lamellae
with the cell boundaries) providing initial pinning centers necessary for
the domain walls to bend and get pressed against the strongly pinning
cell boundaries. However, the estimates obtained in Section~\ref{smco}
suggest that the set of conditions for the realization of this regime
is very strict.

In polytwinned CoPt-type magnets the competition between the anisotropy of
exchange stiffness and magnetostatic interaction controls the orientation of
domain wall segments and the preferential global macrodomain wall orientation.
In particular, the domain wall segments gradually rotate toward the tetragonal
axis during the microstructural coarsening as the $c$-domains become thicker.
If the antiphase boundaries have a preferential orientation, the coercivity
achieves its maximum at the $c$-domain thickness when the domain wall segments
become parallel to the antiphase boundaries.
The same competition also leads to the dynamic rotation of the segments
of a macrodomain wall which is being split by the external field. 
This rotation may result in a dynamic self-locking of domain wall
segments at antiphase boundaries.

The effects discussed in this paper demonstrate that
the competition between exchange stiffness anisotropy and magnetostatic energy
is a crucial driving force behind the formation and dynamics of nanoscale
domain structures in hard magnets.

\begin{acknowledgments}
I am much indebted to V.P. Antropov for numerous
useful discussions and suggestions, and to M. van Schilfgaarde
for the use of his computer codes. I also thank R. Sabirianov and R. Skomski
for useful discussions. A large part of this work was carried out at Ames Laboratory,
which is operated for the U.S. Department of Energy by Iowa State University under
Contract No. W-7405-82. This work was supported by the Director for Energy
Research, Office of Basic Energy Sciences of the U.S. Department of
Energy.
\end{acknowledgments}


\begin{thebibliography}{9}
\bibitem{Bloch-Neel} Simultaneously
the magnetic structure of each domain wall segment transforms from Bloch-type to
Ne\'el-type in the center of the $c$-domain, while it remains Bloch-type in the vicinity
of the twin boundary.
The configuration of the domain wall segments at the initial stage of this
transition for $d\sim20$~nm was shown in Fig.~1 of Ref.~\onlinecite{BA-jap}.
\end{thebibliography}
\end{document}